# TITLE PAGE

**Title:** Diagnostic performance of echocardiography in detecting and differentiating cardiac amyloidosis: a systematic review and meta-analysis.

**Authors:**

Zihang Zhang*[1], Yunjie Chen*[1], Yuanzhou Cao[2], Xinyi Xie[1], Kangming Ji[1], Chuang Yang†[1], Lijun Qian†[1]

**Affiliation:**

[1] Department of Geriatrics, The First Affiliated Hospital with Nanjing Medical University, Nanjing 210029, China;

[2] School of Mathematical Sciences, Nanjing Normal University, Nanjing 210029, China

**Correspondence to:** Lijun Qian, lijunqian@njmu.edu.cn; Chuang Yang, yang@njmu.edu.cn

**\* Drs Zhang and Chen are co-first authors.**

**Running title:**

Differentiating Cardiac Amyloidosis with Echocardiography

**Contributions:**

(I)Conception and design: Lijun Qian; (II) Administrative support: Lijun Qian; (III) Provision of study materials or patients: All authors; (IV) Collection and assembly of data: Zihang Zhang, Yunjie Chen; (V) Data process and analysis: Yuanzhou Cao, Zihang Zhang, Yunjie Chen; (VI) Data interpretation: Zihang Zhang, Yunjie Chen; (VII) Manuscript writing: All authors; (VIII) Final approval of manuscript: All authors.

# Diagnostic performance of echocardiography in detecting and differentiating cardiac amyloidosis: a meta-analysis

**Abstract**

**Aims:** This meta-analysis aimed to evaluate the diagnostic performance of echocardiographic parameters for cardiac amyloidosis (CA), with a focus on subtype stratification and comparisons with healthy controls.

**Methods and Results:** A comprehensive search identified 26 studies published before February 2025, encompassing 3,802 patients. Compared to healthy individuals, CA patients demonstrated significant echocardiographic abnormalities, including reduced left ventricular ejection fraction (LVEF; WMD = -10.65, 95% CI: [-11.84, -9.46]), increased left atrial volume index (WMD = +15.87, 95% CI: [14.35, 17.38]), and thickened posterior wall (WMD = +5.14, 95% CI: [4.85, 5.42]). Subtype analyses revealed that transthyretin cardiac amyloidosis (ATTR-CA) was associated with more pronounced systolic dysfunction than light-chain cardiac amyloidosis (AL-CA), evidenced by lower global longitudinal strain (WMD = -2.02, 95% CI: [-2.66, -1.37]), reduced LVEF (WMD = -5.31, 95% CI: [-6.63, -3.99]), and diminished tricuspid annular plane systolic excursion (WMD = -1.59, 95% CI: [-2.23, -0.95]). Additionally, ATTR-CA patients exhibited greater ventricular wall thickening in both posterior wall (WMD = +1.87, 95% CI: [1.51, 2.23]) and interventricular septum (WMD = +2.24, 95% CI: [1.85, 2.63]).

**Conclusion:** Echocardiography plays a pivotal role in diagnosing CA and distinguishing between AL-CA and ATTR-CA. Key indices such as LVEF and global longitudinal strain are especially valuable for early detection, while subtype-specific patterns highlight distinct underlying pathophysiologies, offering guidance for tailored diagnostic and therapeutic strategies.

**Keywords:** cardiac amyloidosis, echocardiography, global longitudinal strain, immunoglobulin light-chain amyloidosis, transthyretin amyloidosis.

**Introduction**

Cardiac amyloidosis (CA) is a rare yet life-threatening cardiovascular disorder, characterised by the extracellular deposition of misfolded amyloid fibrils within cardiac tissue (1). This pathological accumulation leads to progressive structural and functional deterioration, increased myocardial stiffness, impaired diastolic function, and ultimately, heart failure (2). Of the more than 30 known amyloid precursor proteins, the two principal subtypes of CA are immunoglobulin light-chain amyloidosis (AL-CA) and transthyretin amyloidosis (ATTR-CA) (3-5). The prognosis of CA remains poor, with a median survival ranging from 6 months to 5 years for AL-CA and 3 to 5 years for ATTR-CA (6-8). Given the aggressive nature and high mortality associated with this condition, early and accurate diagnosis is imperative to facilitate timely intervention, slow disease progression, and improve patient outcomes.

Echocardiography, a widely utilised non-invasive imaging modality, plays a pivotal role in assessing cardiac structure and function, diagnosing suspected CA, and guiding long-term disease management (9). Advanced echocardiographic techniques, such as myocardial strain imaging and tissue Doppler imaging, have been explored as potential diagnostic markers for CA. These modalities allow for the evaluation of myocardial deformation and systolic function, which are often impaired in CA patients (10, 11). However, despite these advancements, the diagnostic performance and reliability of specific echocardiographic parameters in differentiating CA from other cardiomyopathies, as well as distinguishing between AL-CA and ATTR-CA, remain uncertain (12).

Although echocardiography provides valuable insights into CA-associated cardiac abnormalities, the lack of consensus on standardized echocardiographic parameters limits its clinical applicability (4, 12). Many studies in this field are constrained by small sample sizes, heterogeneity in methodology, and variations in diagnostic criteria, thereby making it challenging to establish definitive conclusions (13). Furthermore, limited research has systematically compared echocardiographic indices between AL-CA and ATTR-CA, impeding the development of subtype-specific diagnostic and prognostic strategies. A comprehensive synthesis of existing data is therefore essential to address this gap.

This meta-analysis aims to systematically compare echocardiographic cardiac function indices between CA patients and healthy individuals, as well as between AL-CA and ATTR-CA subtypes. The study seeks to provide evidence-based insights into the diagnostic value of echocardiographic markers in CA. By identifying key echocardiographic differences between AL-CA and ATTR-CA, this meta-analysis will contribute to improving early detection, refining risk stratification, and facilitating personalized disease management. Additionally, these findings may inform future research and contribute to the optimization of echocardiographic protocols, thereby enhancing diagnostic accuracy and clinical decision-making.

**Methods**

This systematic review and meta-analysis were conducted in accordance with the Preferred Reporting Items for Systematic Reviews and Meta-Analyses (PRISMA) guidelines (14). As this study was based on publicly available data, approval from an institutional review board was not required.

*Study Selection*

A comprehensive literature search was conducted across PubMed, Embase, and the Cochrane Library to identify relevant studies. The following search terms and keywords were used: "cardiac amyloidosis", "light chain", "amyloid transthyretin", "CA", "AL", "ATTR", "echocardiography", "left ventricular ejection fraction(LVEF)", "global longitudinal strain(GLS)", "left atrial volume index(LAVI)", and "tricuspid annular plane systolic excursion(TAPSE)". Original research articles published up to February 2025 that reported echocardiographic findings in CA patients were included.

*Selection Criteria*

A total of 758 potentially relevant articles were identified through the initial database search. Following a rigorous screening process, 26 studies met the eligibility criteria and were included in the final analysis. Studies were considered eligible if they met the following inclusion criteria: (i) Reported complete study data, (ii) Investigated echocardiographic parameters in cardiac amyloidosis, (iii) Exclusively studied CA patients without other cardiomyopathies, (iv) Were published in English with full-text availability, (v) Included either healthy controls vs. CA patients or comparisons between AL-CA and ATTR-CA patients.

Studies were excluded if they met any of the following exclusion criteria: (i) Contained incomplete or missing data, (ii) Utilized non-echocardiographic imaging modalities (e.g., cardiac magnetic resonance imaging), (iii) Included patients with additional coexisting cardiac diseases, (iv) Were review articles or case reports, (v) Contained overlapping patient cohorts (e.g., studies with similar recruitment intervals from the same institution).

*Data Collection*

Two independent researchers assessed study eligibility using a standardized data extraction form, with a third reviewer verifying the data for consistency and accuracy. The following key study characteristics were extracted: Study name and design, Echocardiographic equipment used, Patient nationality, Number of patients and their mean age, Type of cardiac amyloidosis (AL-CA or ATTR-CA), and Major echocardiographic findings. All echocardiographic parameters were analyzed as continuous variables. For each included study, the mean, standard deviation, and sample size were extracted for both the experimental and control groups.

*Quality Assessment*

To assess the methodological quality of the included studies, two researchers independently evaluated the risk of bias using the Cochrane Collaboration's tool (RevMan V5.4) (15). The following six domains of bias were systematically assessed: Selection bias: Random sequence generation and allocation concealment, Performance bias: Blinding of participants and study personnel, Detection bias: Blinding of outcome assessment, Attrition bias: Incomplete outcome data, and Reporting bias: Selective reporting. Each study was assigned a rating of "high," "low," or "unclear" risk of bias, with discrepancies resolved through discussion with the corresponding author. Bias risk plots and summary tables were generated to provide a visual representation of study quality.

*Statistical Analysis*

Meta-analysis was conducted using RevMan V5.4, applying a fixed-effect model to synthesize data across studies. Heterogeneity was assessed using the Cochrane Q test and the $I^2$ statistic, with the chi-square test employed to quantify its magnitude. If substantial heterogeneity was detected, potential sources were explored, and results were discussed accordingly. To evaluate publication bias, Begg's funnel plot visual inspection and Egger's regression test were performed. A p-value <0.05 was considered to indicate statistical significance.

**Results**

*Study Characteristics*

A total of 758 articles were identified through the literature search (Figure 1). Following the removal of 98 duplicates and the exclusion of 261 review articles or case reports, 235 articles were deemed irrelevant to the study focus. Among the remaining 134 articles, 18 were excluded due to incomplete data, 19 for utilizing non-echocardiographic imaging modalities including cardiac magnetic resonance imaging （CMR), 42 for including patients with coexisting cardiac diseases, 19 due to the unavailability of full-text access, and 10 for potential patient overlap (similar recruitment intervals from the same location). Ultimately, 26 studies met the inclusion criteria and were incorporated into the final analysis (16-41).

The fundamental study characteristics are summarized in Tables 1 and 2. Eighteen studies provided echocardiographic data comparing CA patients with healthy controls, whereas nine studies included comparisons between AL-CA and ATTR-CA patients (with one study being duplicated). The risk of bias across these studies is illustrated in Figure 2, while Figures 3–5 depict the comparative analysis between CA patients and healthy controls, and Figures 6–7 compare echocardiographic parameters between AL-CA and ATTR-CA.

*Impact of CA on Left Ventricular Systolic Function*

To evaluate differences in left ventricular systolic function between CA patients and healthy controls, LVEF and GLS were assessed. LVEF, a fundamental indicator of cardiac function, demonstrated significant predictive value for CA, with a weighted mean difference (WMD) of 10.65 (95% CI: [9.46, 11.84]; $P < 0.00001$) and low heterogeneity ($I^2 = 29\%$). GLS, derived from speckle-tracking echocardiography, also exhibited predictive significance but yielded a lower pooled effect size (WMD = -6.94; 95% CI: [-7.32, -6.56]; $P < 0.00001$) with substantial heterogeneity ($I^2 = 86\%$). The funnel plot analysis indicated no significant publication bias in this study.

*Impact of CA on Left Ventricular Diastolic Function*

To examine differences in left ventricular diastolic function, the following echocardiographic indices were analysed: E/A ratio, E/e' ratio, and LAVI. E/A ratio

(WMD = -0.56; 95% CI: [-0.62, -0.50]; P < 0.00001), E/e' ratio (WMD = -9.28; 95% CI: [-9.83, -8.72]; P < 0.00001) and LAVI (WMD = -15.87; 95% CI: [-17.38, -14.35]; P < 0.00001) were achieved. Among these indices, the E/e' ratio demonstrated the largest effect size (WMD = -9.28), with moderate heterogeneity ($I^2 = 48\%$), indicating high clinical reliability as a diagnostic marker for CA. Although LAVI exhibited the greatest absolute effect size, it was associated with higher heterogeneity ($I^2 = 87\%$). The funnel plot revealed no significant publication bias.

*Impact of CA on Wall Thickness and Chamber Size*

To explore differences in wall thickness and chamber size (42), the following echocardiographic parameters were analyzed: posterior wall thickness (PWT), interventricular septal thickness at end-diastole (IVSD), and left ventricular end-systolic volume (ESV). CA patients exhibited significant posterior wall thickening, with a weighted mean difference ranging from -5.42 to -4.85 (P < 0.00001) and low heterogeneity ($I^2 = 25\%$), supporting the robustness of this finding. IVSD demonstrated higher heterogeneity ($I^2 = 81\%$, P<0.00001), but sensitivity analysis confirmed a statistically significant difference (WMD= -4.69; 95% CI: [-4.95, -4.43]; P < 0.00001). Conversely, the pooled effect size for ESV was not statistically significant (WMD = -0.70; 95% CI: [-3.19, 1.79]; P = 0.58), suggesting no significant difference in chamber size between CA patients and healthy controls. Funnel plot and Egger's test (P = 0.19) confirmed the absence of publication bias.

*Impact of AL-CA and ATTR-CA on Cardiac Systolic Function*

To compare cardiac systolic function between AL-CA and ATTR-CA subtypes, a systematic analysis was conducted using: GLS, LVEF, and TAPSE. AL-CA patients exhibited higher absolute GLS values compared to ATTR-CA (WMD= -2.02; 95% CI: [-2.66, -1.37]; P < 0.00001) with moderate heterogeneity ($I^2 = 72\%$). LVEF was significantly higher in AL-CA than in ATTR-CA (WMD = 5.31; 95% CI: [3.99, 6.63]; P < 0.00001, $I^2 = 0\%$). TAPSE values were also higher in AL-CA (WMD = 1.59; 95% CI: [0.95, 2.23]; P = 0.38).

These findings suggest that AL-CA patients included in this study demonstrated better myocardial deformation capacity and superior systolic function in both ventricles. This discrepancy may be attributable to the early-stage nature of the study population, where myocardial function in AL-CA had not yet deteriorated significantly. The funnel plot revealed no evidence of publication bias, further supporting the validity of these findings.

*Differences in Wall Thickness Between AL-CA and ATTR-CA*

To assess ventricular wall thickness differences between AL-CA and ATTR-CA, two key parameters were evaluated: IVSD and PWT. Compared to AL-CA patients, ATTR-CA patients exhibited significantly higher IVSD values (WMD = -2.24; 95% CI: [-2.63, -1.85]; P < 0.00001; $I^2 = 61\%$) despite higher heterogeneity. Similarly, PWT was significantly greater in ATTR-CA (WMD = -1.87; 95% CI: [-2.23, -1.51]; P < 0.00001; $I^2 = 9\%$).

These results indicate that ATTR-CA patients exhibit more pronounced structural changes, with greater amyloid deposition and higher susceptibility to myocardial hypertrophy, leading to more severe myocardial dysfunction. This aligns with the observed differences in systolic function, further supporting the conclusion that ATTR-CA results in more extensive cardiac involvement than AL-CA. The funnel plots revealed no evidence of publication bias, reinforcing the reliability of the findings.

**Discussion**

This meta-analysis comprehensively evaluated the diagnostic performance of echocardiography in detecting and differentiating cardiac amyloidosis (CA), focusing on comparisons between AL-CA and ATTR-CA. The results underscore that echocardiographic parameters—including LVEF, GLS，PWT—differ significantly between CA patients and healthy controls. In particular, LVEF showed strong diagnostic value with relatively low heterogeneity, supporting its utility as a clinically relevant marker. Additionally, ATTR-CA patients showed more advanced wall thickening and greater cardiac dysfunction compared to AL-CA patients, suggesting meaningful structural and functional differences between the subtypes.

Echocardiography's ability to detect these differences highlights its pivotal role in early diagnosis and subtype differentiation. The marked reductions in LVEF and GLS among CA patients can be attributed to differing pathophysiology: AL-CA involves deposition of immunoglobulin light chains, often resulting in diffuse myocardial infiltration, while ATTR-CA is characterized by transthyretin-derived fibrils causing pronounced hypertrophy and diastolic dysfunction (43–46). These findings reflect established knowledge on amyloid pathology and emphasize the capacity of echocardiography to provide pathophysiological insights that inform clinical decision-making.

Interestingly, AL-CA patients demonstrated better systolic function and myocardial deformation than ATTR-CA patients in this analysis. This may be due to earlier diagnosis in AL-CA, as AL amyloidosis often presents more acutely and leads to earlier medical attention. In contrast, ATTR-CA progresses more insidiously, often resulting in greater myocardial damage by the time of diagnosis. These temporal differences may explain the apparent contradiction with studies reporting more severe outcomes in AL-CA patients, highlighting the importance of considering disease stage in echocardiographic interpretation (46, 47).

Previous research supports the role of echocardiography in detecting CA-related myocardial involvement and distinguishing it from other cardiomyopathies (48, 49). This study builds on prior work by directly comparing echocardiographic parameters between CA subtypes and healthy individuals. Consistent with prior findings, key markers such as LVEF and GLS emerged as reliable indicators of myocardial dysfunction. PWT also proved to be a valuable structural marker, differentiating CA from hypertrophic cardiomyopathy and hypertensive heart disease (50). Notably, while amyloid cardiomyopathy often preserves end-systolic volume (ESV), it manifests through myocardial thickening and dysfunction—further reinforcing the diagnostic importance of wall thickness and functional parameters (49, 51).

Contrary to some reports suggesting AL-CA has more severe clinical consequences (52), this analysis found that ATTR-CA patients had worse systolic function and greater structural abnormalities. This discrepancy may stem from variation in patient populations, inclusion criteria, or disease stages across studies (47). It is also important to consider subtype heterogeneity within ATTR-CA (53). For instance, mutations such as V112I are associated with more severe cardiac dysfunction (54), while V30M subtype expression varies by phenotype—type A presenting with more advanced cardiomyopathy and type B with milder cardiac involvement (55).

Clinically, these findings support the use of echocardiographic parameters for early identification and subtype differentiation in CA. Parameters such as LVEF and GLS can assist in recognizing CA before the onset of overt heart failure, allowing for earlier interventions. Moreover, echocardiographic differences between AL-CA and ATTR-CA may help guide tailored treatment strategies—such as chemotherapy for AL-CA or TTR stabilizers for ATTR-CA—by providing insights into disease burden and progression. As the understanding of amyloidosis evolves, echocardiography may also help monitor therapeutic response and disease trajectory.

This study also offers theoretical contributions by reinforcing distinct pathophysiological mechanisms underlying the two CA subtypes. The systematic meta-analytic approach provides a high-level synthesis of existing evidence, supporting the potential for echocardiography to serve not only as a diagnostic modality but also as a surrogate marker of disease severity.

Nevertheless, some limitations must be acknowledged. The relatively small number of included studies, along with variation in echocardiographic techniques and patient selection criteria, introduces heterogeneity that may affect generalizability. The predominance of early-stage AL-CA patients in the analyzed cohorts may underrepresent the full disease spectrum. Moreover, as most included studies were observational and cross-sectional, causal relationships cannot be established. Potential confounding variables and selection bias may also influence the findings.

Future research should expand on this foundation by incorporating longitudinal data to assess changes in echocardiographic parameters over time. Larger, more diverse cohorts including late-stage patients will help clarify how echocardiographic features evolve with disease progression. Standardized echocardiographic protocols for CA assessment are also needed to enhance reproducibility and clinical application. Finally, advanced imaging modalities such as 3D echocardiography and novel strain imaging techniques may offer additional diagnostic and prognostic value.

**Conclusion**

This meta-analysis confirms the value of echocardiography in diagnosing and differentiating cardiac amyloidosis, highlighting LVEF and GLS as key markers and revealing distinct structural and functional patterns between AL-CA and ATTR-CA subtypes.

**Disclosure**

The authors have no conflicts of interest to disclose.

**Funding**

This study was supported by the educational research project of Nanjing Medical University (2023YJS-LX011 and 2023YJS-LX012), National Natural Science Foundation of China (82102073 and 82203216), Natural Science Foundation of Jiangsu Province (BK20220724) and the Postdoctoral Research Fund of Nanjing Municipality(2024BHS209).

**Acknowledgements**

We thank the study patients and investigators and the sponsors for their contributions to data collection and analysis, assistance with statistical analysis, or critical review of the manuscript. We thank figdraw for providing for us to draw pictures.

J Pathol. 2008;216(2):253-61.

## Figure Legends

**Figure 1.** PRISMA flowchart.

**Figure 2**. Quality assessment of the included studies. A systematic evaluation of bias risk across the selected studies, assessed using the Cochrane Collaboration's tool.

**Figure 3.** Forest plots (A, B) and funnel plots (C, D) illustrating the differences in LVEF (A, C) and GLS (B, D) between CA patients and healthy controls.

**Figure 4**. Forest plots (A, B, C) and funnel plots (D, E, F) depicting the differences in E/A (A, D), E/e' (B, E), and LAVI (C, F) in the comparison between CA patients and healthy controls.

**Figure 5.** Forest plots (A, B, C) and funnel plots (D, E, F) showing the differences in PWT (A, D), IVSD (B, E), and ESV (C, F) between CA patients and healthy controls.

**Figure 6.** Forest plots (A, B, C) and funnel plots (D, E, F) illustrating the differences in GLS (A, D), LVEF (B, E), and TAPSE (C, F) between AL-CA and ATTR-CA patients.

**Figure 7.** Forest plots (A, B) and funnel plots (C, D) demonstrating the differences in IVSD (A, C) and PWT (B, D) between AL-CA and ATTR-CA patients, highlighting structural variations between the two subtypes.

**Table 1. Baseline characteristics among studies with CA and control patients.**

| First author/s, year | Study design | Country of origin | Echocardiographic device | Number of patients with CA and control group | Mean age of patients with CA and control group | Kinds of CA | Main echocardiographic findings |
|---|---|---|---|---|---|---|---|
| Queenie et al, 2016 | M, R | Australia | Vivid 7 GE, iE33 Philips | 46 / 46 | 60±12 / 59±13 | AL, ATTR | LV-GLS, GCS, E/e', LVMI, RWT |
| Huang et al, 2024 | S, R | China | vivid E9 or E95 ultrasound | 63 / 33 | 59±7 / 54±11 | AL | LVEF, LAVI, E/e', PWT |
| Vincenzo et al, 2016 | S, P | America | Vivid 7 ultra-sound scanner | 33 / 30 | 62±11.9 / 57.7±11.8 | AL | LVEF, LV-GLS, E/e', LVMI, PWT, LAVI |
| Vanessa et al, 2019 | S, P | America | iE33 ultrasound system | 47 / 24 | 63 ± 11 / 61 ± 13 | AL | LVEF, LV-GLS, LAVI |
| Juan et al, 2024 | M, R | 4 countries | Philips or GE | 544 / 174 | 73 ± 11 / 74 ± 12 | AL, ATTR | LVEF, LV-GLS, E/e', LVMI, PWT, LAVI |
| Yoshihito et al, 2022 | S, P | America | GE Vivid7 or Vivid9 | 18 / 16 | 71 ± 9 / 43 ± 18 | NA | LVEF, LAVI, E/e', LV-GLS, EDV, ESV |
| Nikola et al, 2022 | R | NA | Vivid E9, GE | 59 / 150 | 72.5 ± 10.2 / 33.8 ± 11.5 | AL, ATTR | LVEF, LVDD, E/e', EDV, ESV, LVMI, LAVI |

25| Study | Design | Country | Equipment | N (CA/Control) | Age (CA/Control) | Type | Measurements |
|---|---|---|---|---|---|---|---|
| Douglas et al, 2021 | S, R | America | Vivid E95 | 71 / 29 | 53.2（14.2）/ 67.9（11.7） | NA | LVEF, GLS |
| Antonio et al, 2018 | S, P | NA | Vivid E9 | 23 / 23 | 62.1 ± 10.6 / 59.1 ± 9.5 | AL | LVEF, LVMI, LV-GLS |
| Attila et al, 2022 | R | Japan | 3D-STE | 27 / 20 | 62.7±9.1 / 59.3±3.8 | AL，ATTR | LVEF, EDV, ESV, E/A, LVDD |
| Alberto et al, 2022 | S, R | Italy | EchoPAC 12.0, GE | 261 / 162 | 78 (72–83) / 79 (74–84) | AL，ATTR | LVEF, LVMI, RWT, E/e', LAVI, LV-GLS |
| Kotaro et al, 2017 | M, R | America | Vivid 7 | 124 / 20 | 64.1 ± 11.9 / 66.5 ± 5.0 | AL，ATTR | LV-GLS, E/e', E/A, LAVI |
| Jeremy et al, 2023 | M, R | America | iE33 | 67 / 45 | 71 ± 13 / 68 ± 5 | AL，ATTR | PWT, GLS, LVEF, E/e', LVMI |
| Jiang et al, 2024 | S, R | China | NA | 213 / 181 | 58（49，64）/ 48（43，54） | NA | PWT, LVEF, E/A |
| Ines et al, 2023 | S, R | NA | E6-GE | 33 / 33 | 68 (62.5–77.5) / 58 (53–65) | ATTR | LVEF, LVMI, E/e', LAVI, GLS |
| Yu et al, 2018 | S, P | China | 3D-STI | 32 / 16 | 63±7 / 53±14 | NA | GLS, E/A |
| Dóra et al, 2017 | R | NA | 3D-STE | 16 /16 | 64.0±9.6 /58.2±7.2 | AL | GLS, E/A, LVEF, GCS |
| Chiharuko et al, 2015 | S, R | Japan | Vivid 7 and E9 | 15 /20 | 62 ± 14 / 58 ± 7 | NA | LAVI, PWT, LVEF, LVMI, E/A, E/e' |

Values are presented as median (interquartile range) or mean ± standard deviation. M, multicentre; S, single centre; P, prospective; R, retrospective; NA, not available; CA, cardiac amyloidosis; AL, light chain; ATTR, amyloid transthyretin; LVEF, left ventricular ejection fraction; LV-GLS, left ventricular-global longitudinal strain; LVMI, left Ventricular Mass Index; LAVI, left Atrial Volume Index; PWT, posterior wall thickness; RWT, relative Wall Thickness; EDV, end diastolic volume; ESV, end systolic volume; 3D-STE, three-dimensional speckle-tracking echocardiography; 3D-STI, three-dimensional speckle-tracking imaging.

**Table 2. Baseline characteristics among studies with immunoglobulin light-chain amyloidosis (AL-CA) and transthyretin amyloidosis (ATTR-CA).**

| First author/s, year | Study design | Country of origin | Echocardiographic device | Number of patients with AL and ATTR cardiac amyloidosis | Mean age of patients with AL and ATTR cardiac amyloidosis | Main echocardiographic findings |
|---|---|---|---|---|---|---|
| Aaisha *et al*, 2024 | M, P | Australia | GE EchoPAC Version 204 | 86 / 86 | 62 (54–72) / 74 (65–80) | LVEF, RWT, TAPSE, LV-GLS |
| Yoshihito *et al*, 2025 | R | Japan | E9 and E95 | 10 / 25 | 65 ± 11 / 78 ± 8 | PWT, IVSD, LVEF |
| Cristiane *et al*, 2024 | S, P | NA | iE33 or EPIQ ultrasound imaging systems | 89 / 131 | 64 (58, 70) / 79 (74, 83) | GLS, PWT, RWT, TAPSE, IVSD, LVEF |
| Brody *et al*, 2024 | S, R | America | Philips, GE, and Siemens | 32 / 53 | 58 ± 10 / 78 ± 8 | GLS, PWT, RWT, IVSD, LVEF |
| Alberto *et al*, 2022 | M, R | Australia | TomTec-Arena version 4.6 | 117 / 144 | 73 (65–78) / 82 (76–85) | GLS, PWT, RWT, TAPSE, IVSD, LVEF |
| Osnat *et al*, 2021 | S, R | Israel | NA | 31 / 36 | 65 (60, 71) / 80 (70, 85) | PWT, RWT, TAPSE, LVEF |
| Nowell *et al*, 2020 | S, R | NA | iE33 | 36 / 57 | 63 ± 11 / 74 ± 8 | PWT, TAPSE, IVSD, LVEF |
| Minako *et al*, 2015 | S, R | Japan | Vivid 7 System | 18 / 28 | 66 ± 8 / 78 ± 6 | IVSD, PWT |
| Francesco *et al*, 2015 | S, R | Italy | Vivid 7 System | 45 / 48 | 68.2 ± 10.2 / 75.8 ± 9.7 | PWT, IVSD, LVEF, TAPSE |

Values are presented as median (interquartile range) or mean ± standard deviation. CA, cardiac amyloidosis; AL, light chain; ATTR, amyloid transthyretin; LVEF, left ventricular ejection fraction; LV-GLS, left ventricular-global longitudinal strain; IVSD, internal ventricular septal diameter; TAPSE, tricuspid annular systolic excursion; PWT, posterior wall thickness; RWT, relative Wall Thickness; M, multicentre; S, single centre; P, prospective; R, retrospective; NA, not available.

# Word count

1. Title:  word = 12

2. Abstract: word = 233

3. Main text: word = 2616

4. References: word = 1603

5. Figure legends: word = 193

6. Tables: word = 877

7. Total word count = 5534

Table 1. Baseline characteristics among studies with CA and control patients.

| First author/s, year | Study design | Country of origin | Echocardiographic device | Number of patients with CA and control group | Mean age of patients with CA and control group | Kinds of CA | Main echocardiographic findings |
|---|---|---|---|---|---|---|---|
| Queenie et al, 2016 | M, R | Australia | Vivid 7 GE, iE33 Philips | 46 / 46 | 60±12 /59±13 | AL，ATTR | LV-GLS, GCS, E/e', LVMI, RWT |
| Huang et al, 2024 | S, R | China | vivid E9 or E95 ultrasound | 63 / 33 | 59±7 / 54±11 | AL | LVEF, LAVI, E/e', PWT |
| Vincenzo et al, 2016 | S, P | America | Vivid 7 ultra-sound scanner | 33 / 30 | 62±11.9 / 57.7±11.8 | AL | LVEF, LV-GLS, E/e', LVMI, PWT, LAVI |
| Vanessa et al, 2019 | S, P | America | iE33 ultrasound system | 47 / 24 | 63 ± 11 / 61 ± 13 | AL | LVEF, LV-GLS, LAVI |
| Juan et al, 2024 | M, R | 4 countries | Philips or GE | 544 / 174 | 73 ± 11 / 74 ± 12 | AL，ATTR | LVEF, LV-GLS, E/e', LVMI, PWT, LAVI |
| Yoshihito et al, 2022 | S, P | America | GE Vivid7 or Vivid9 | 18 / 16 | 71 ± 9 / 43 ± 18 | NA | LVEF, LAVI, E/e', LV-GLS, EDV, ESV |
| Nikola et al, 2022 | R | NA | Vivid E9, GE | 59 / 150 | 72.5 ± 10.2 / 33.8 ± 11.5 | AL，ATTR | LVEF, LVDD, E/e', EDV, ESV, LVMI, LAVI |
| Douglas et al, 2021 | S, R | America | Vivid E95 | 71 / 29 | 53.2（14.2）/ 67.9（11.7） | NA | LVEF, GLS |
| Antonio et al, 2018 | S, P | NA | Vivid E9 | 23 / 23 | 62.1 ± 10.6 / 59.1 ± 9.5 | AL | LVEF, LVMI, LV-GLS |
| Attila et al, 2022 | R | Japan | 3D-STE | 27 / 20 | 62.7±9.1 / 59.3±3.8 | AL，ATTR | LVEF, EDV, ESV, E/A, LVDD |
| Alberto et al, 2022 | S, R | Italy | EchoPAC 12.0, GE | 261 / 162 | 78 (72–83) / 79 (74–84) | AL，ATTR | LVEF, LVMI, RWT, E/e', LAVI, LV-GLS |
| Kotaro et al, 2017 | M, R | America | Vivid 7 | 124 / 20 | 64.1 ± 11.9 / 66.5 ± 5.0 | AL，ATTR | LV-GLS, E/e', E/A, LAVI |
| Jeremy et al, 2023 | M, R | America | iE33 | 67 / 45 | 71 ± 13 / 68 ± 5 | AL，ATTR | PWT, GLS, LVEF, E/e', LVMI |
| Jiang et al, 2024 | S, R | China | NA | 213 / 181 | 58（49，64）/ 48（43，54） | NA | PWT, LVEF, E/A |
| Ines et al, 2023 | S, R | NA | E6-GE | 33 / 33 | 68 (62.5–77.5) / 58 (53–65) | ATTR | LVEF, LVMI, E/e', LAVI, GLS |
| Yu et al, 2018 | S, P | China | 3D-STI | 32 / 16 | 63±7 / 53±14 | NA | GLS, E/A |
| Dóra et al, 2017 | R | NA | 3D-STE | 16 /16 | 64.0±9.6 /58.2±7.2 | AL | GLS, E/A, LVEF, GCS |
| Chiharuko et al, 2015 | S, R | Japan | Vivid 7 and E9 | 15 /20 | 62 ± 14 / 58 ± 7 | NA | LAVI, PWT, LVEF, LVMI, E/A, E/e' |

Values are presented as median (interquartile range) or mean ± standard deviation. M, multicentre; S, single centre; P, prospective; R, retrospective; NA, not available; CA, cardiac amyloidosis; AL, light chain; ATTR, amyloid transthyretin; LVEF, left ventricular ejection fraction; LV-GLS, left ventricular-global longitudinal strain; LVMI, left Ventricular Mass Index; LAVI, left Atrial Volume Index; PWT, posterior wall thickness; RWT, relative Wall Thickness; EDV, end diastolic volume; ESV, end systolic volume; 3D-STE, three-dimensional speckle-tracking echocardiography; 3D-STI, three-dimensional speckle-tracking imaging.

Table 2

Table 2. Baseline characteristics among studies with immunoglobulin light-chain amyloidosis (AL-CA) and transthyretin amyloidosis (ATTR-CA).

| First author/s, year | Study design | Country of origin | Echocardiographic device | Number of patients with AL and ATTR cardiac amyloidosis | Mean age of patients with AL and ATTR cardiac amyloidosis | Main echocardiographic findings |
|---|---|---|---|---|---|---|
| Aaisha *et al*, 2024 | M, P | Australia | GE EchoPAC Version 204 | 86 / 86 | 62 (54–72) / 74 (65–80) | LVEF, RWT, TAPSE, LV-GLS |
| Yoshihito *et al*, 2025 | R | Japan | E9 and E95 | 10 / 25 | 65 ± 11 / 78 ± 8 | PWT, IVSD, LVEF |
| Cristiane *et al*, 2024 | S, P | NA | iE33 or EPIQ ultrasound imaging systems | 89 / 131 | 64 (58, 70) / 79 (74, 83) | GLS, PWT, RWT, TAPSE, IVSD, LVEF |
| Brody *et al*, 2024 | S, R | America | Philips, GE, and Siemens | 32 / 53 | 58 ± 10 / 78 ± 8 | GLS, PWT, RWT, IVSD, LVEF |
| Alberto *et al*, 2022 | M, R | Australia | TomTec-Arena version 4.6 | 117 /144 | 73 (65–78) / 82 (76–85) | GLS, PWT, RWT, TAPSE, IVSD, LVEF |
| Osnat *et al*, 2021 | S, R | Israel | NA | 31 / 36 | 65 (60, 71) / 80 (70, 85) | PWT, RWT, TAPSE, LVEF |
| Nowell *et al*, 2020 | S, R | NA | iE33 | 36 / 57 | 63 ± 11 / 74 ± 8 | PWT, TAPSE, IVSD, LVEF |
| Minako *et al*, 2015 | S, R | Japan | Vivid 7 System | 18 / 28 | 66 ± 8 / 78 ± 6 | IVSD, PWT |
| Francesco *et al*, 2015 | S, R | Italy | Vivid 7 System | 45 / 48 | 68.2 ± 10.2 / 75.8 ± 9.7 | PWT, IVSD, LVEF, TAPSE |

Values are presented as median (interquartile range) or mean ± standard deviation. CA,cardiac amyloidosis; AL, light chain; ATTR, amyloid transthyretin; LVEF, left ventricular ejection fraction; LV-GLS, left ventricular-global longitudinal strain; IVSD, internal ventricular septal diameter ; TAPSE,tricuspid annular systolic excursion; PWT, posterior wall thickness; RWT, relative Wall Thickness; M, multicentre; S, single centre; P, prospective; R, retrospective; NA, not available.

Figure 1

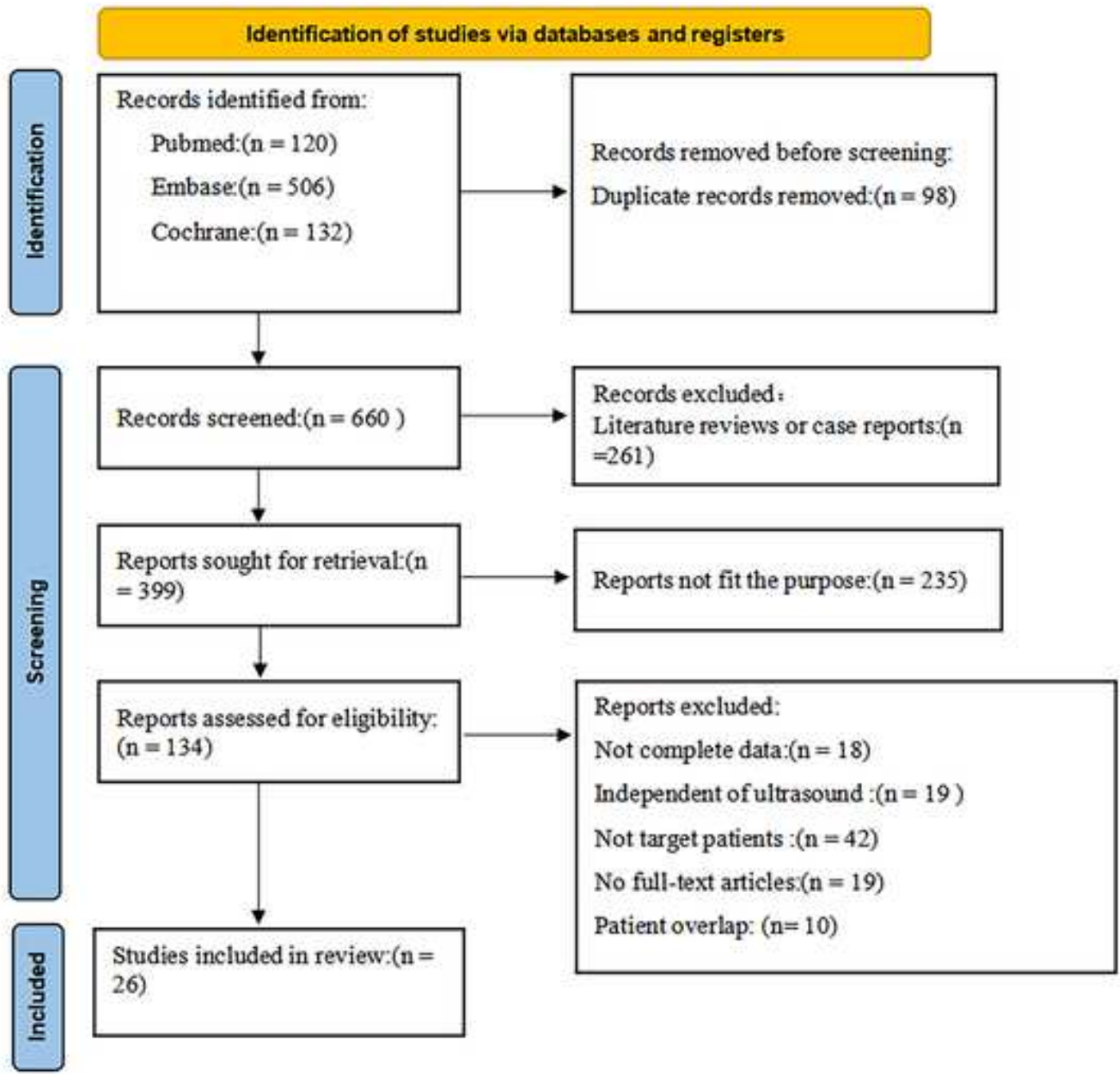

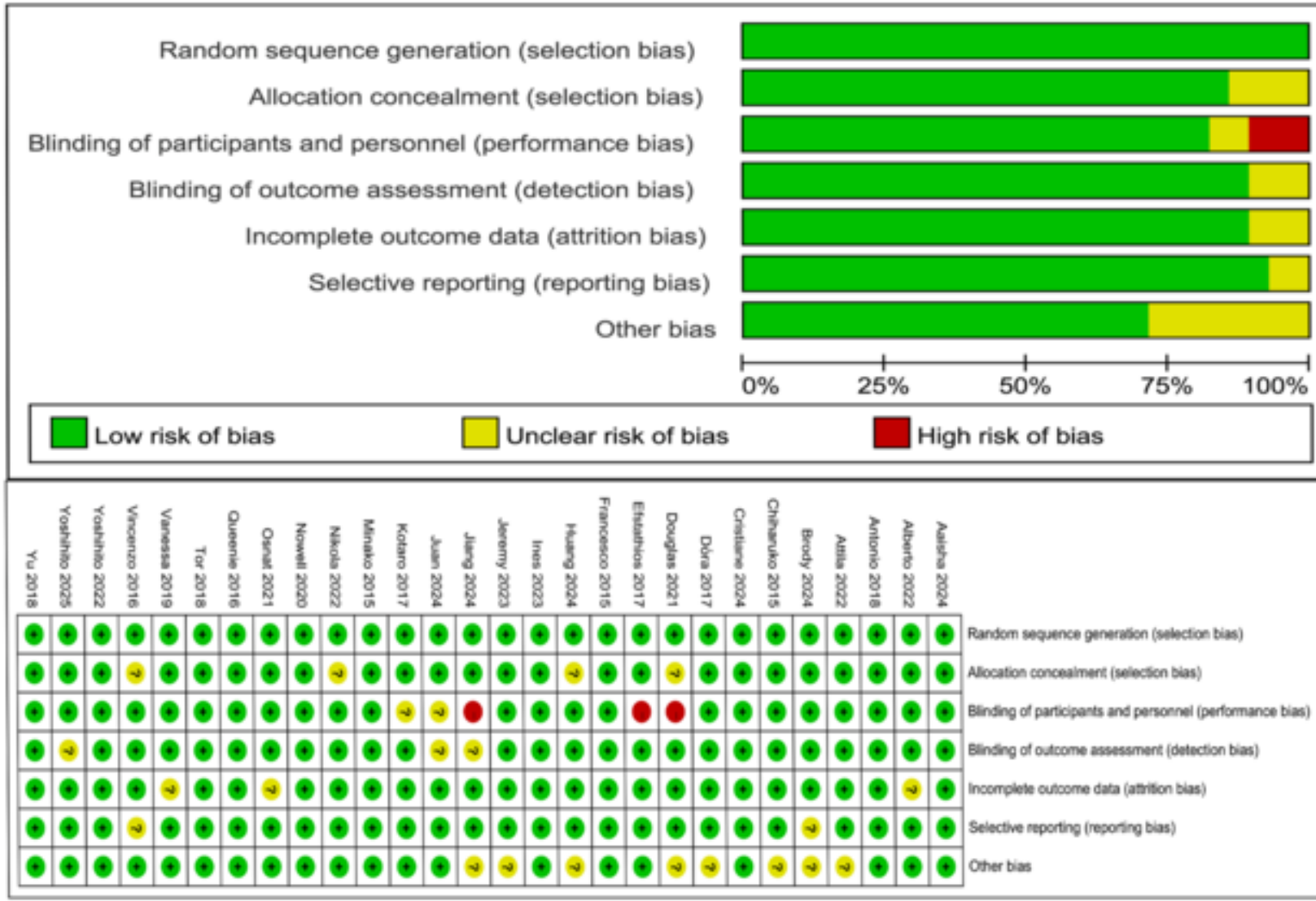

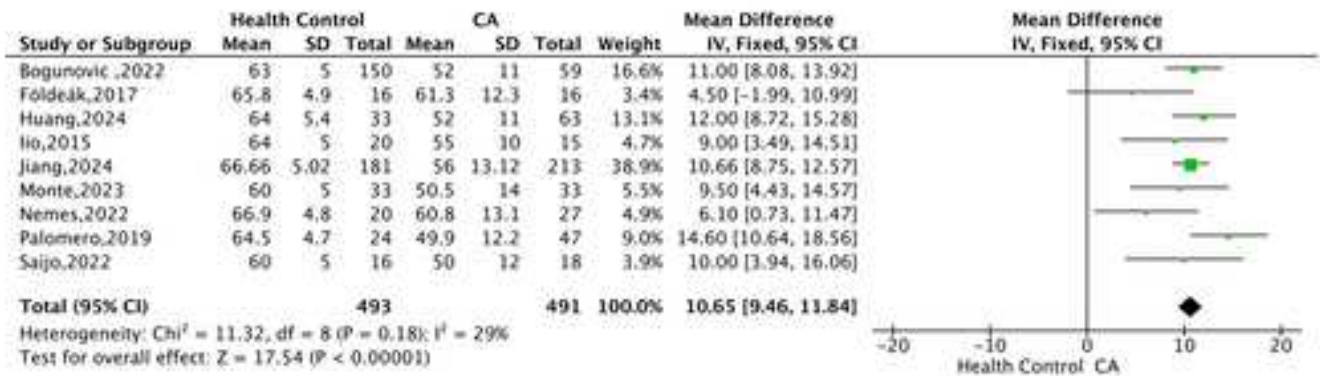
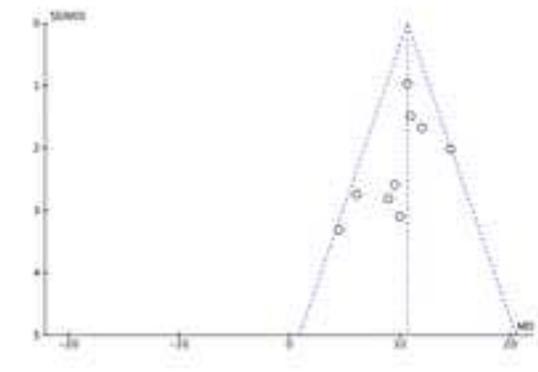
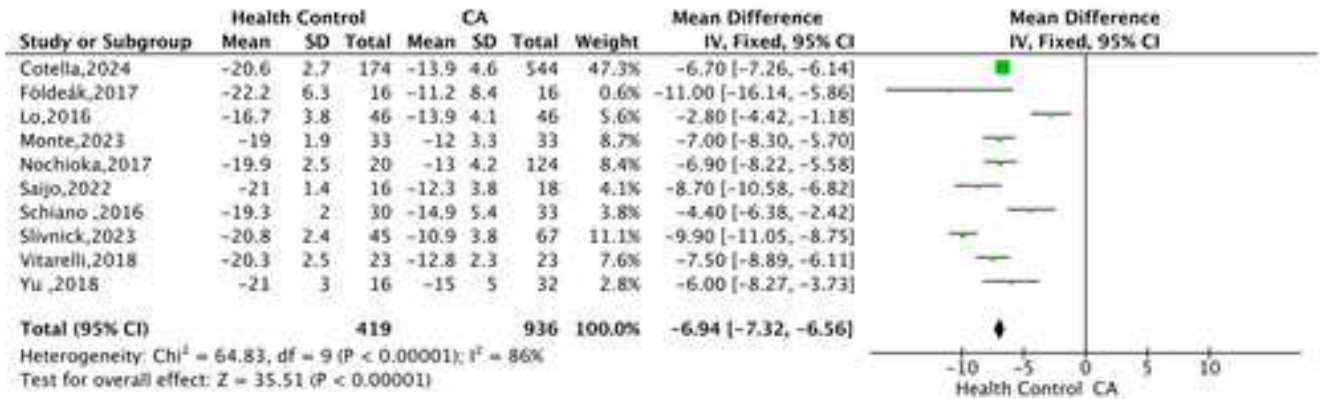
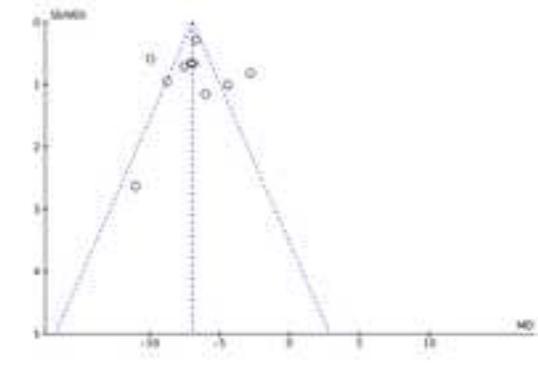

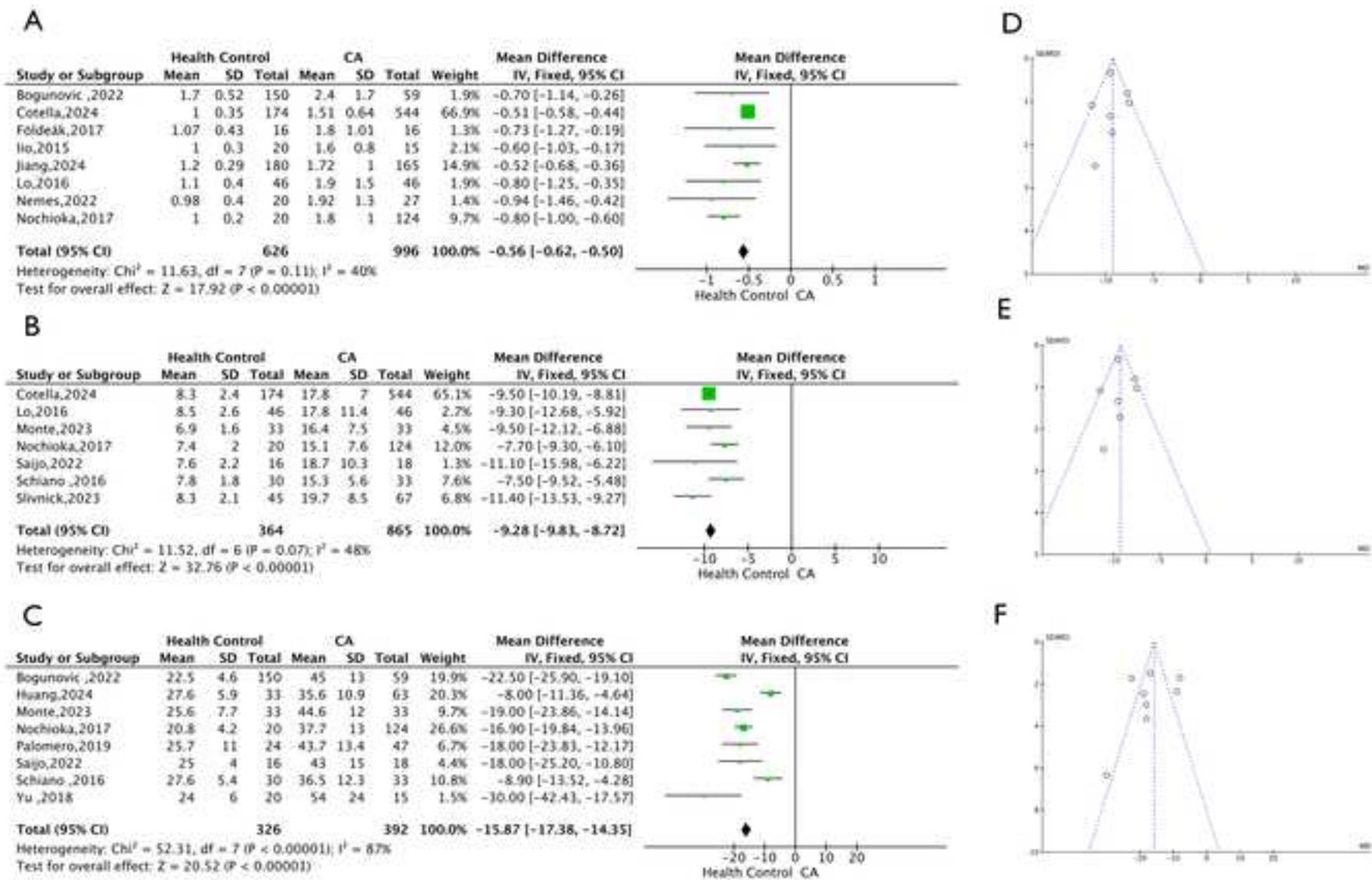

A
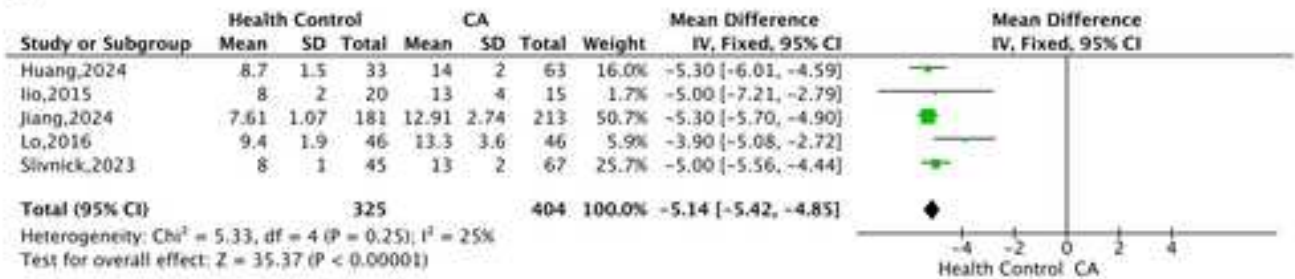
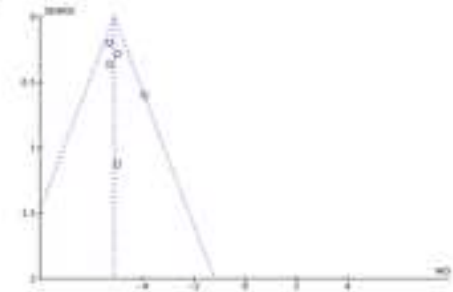
D

B
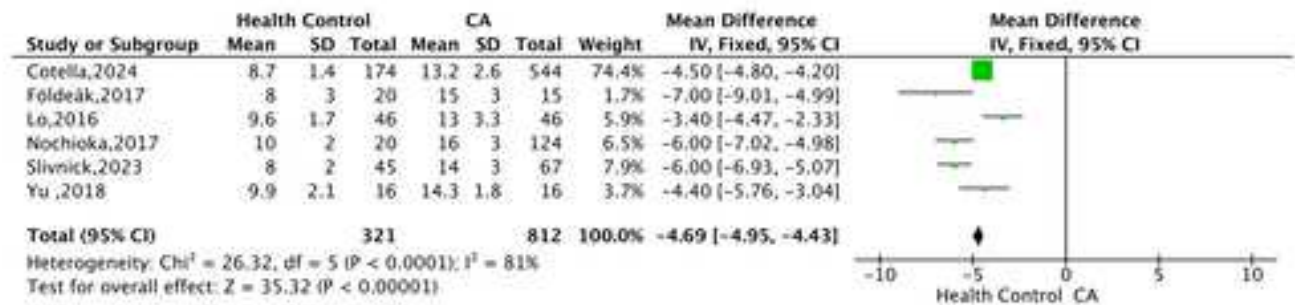
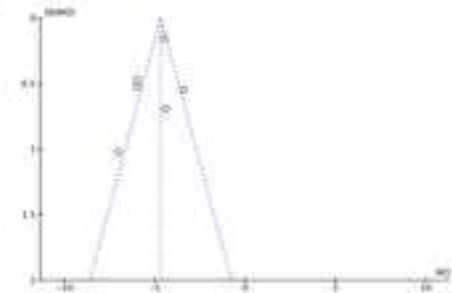
E

C
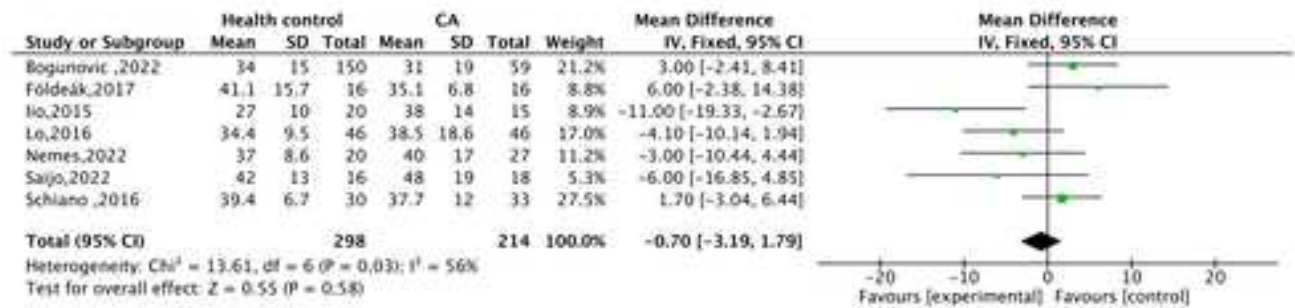
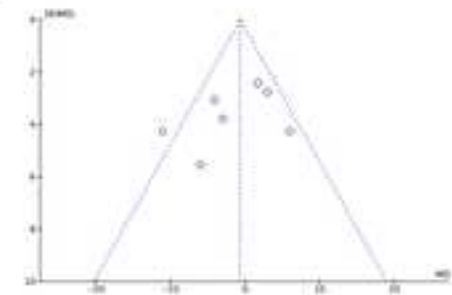
F

Figure 6

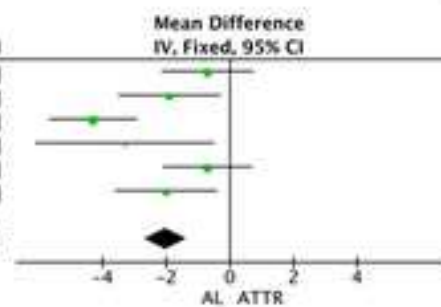
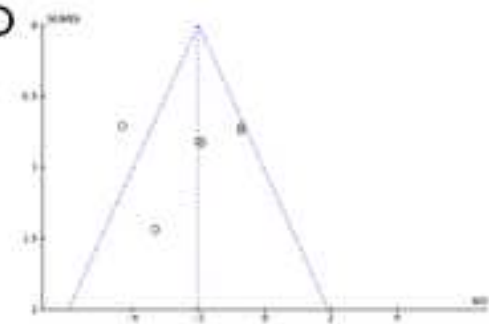
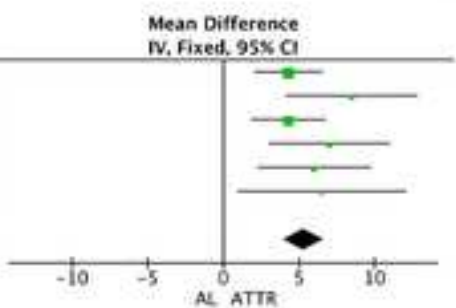
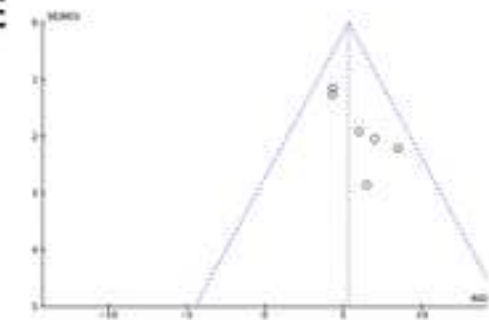
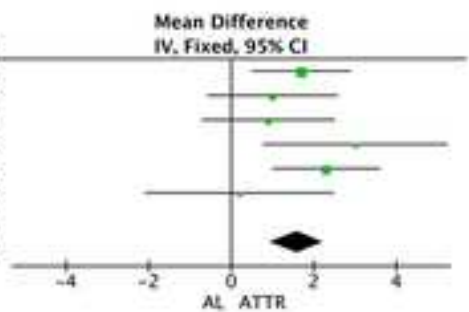
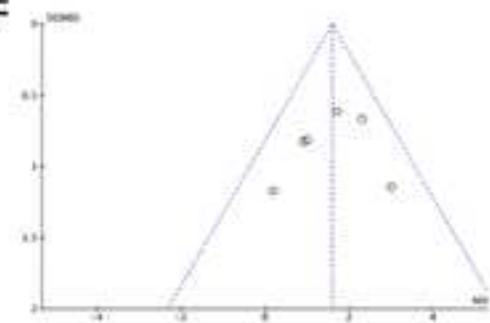

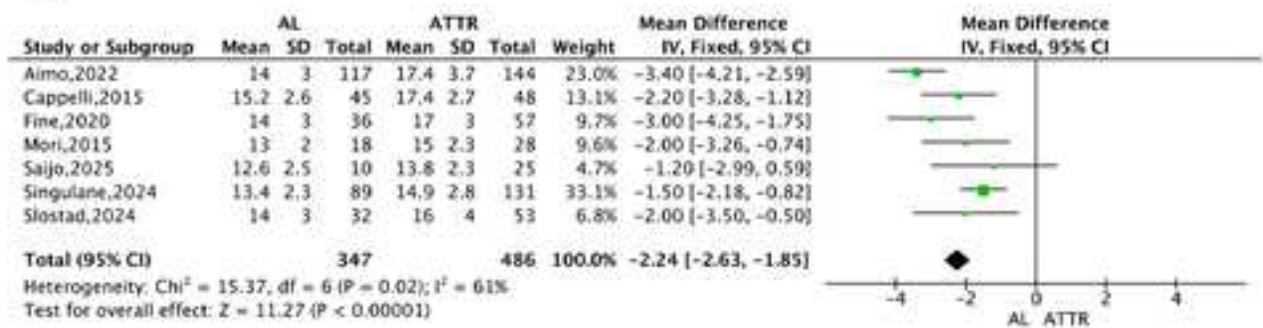

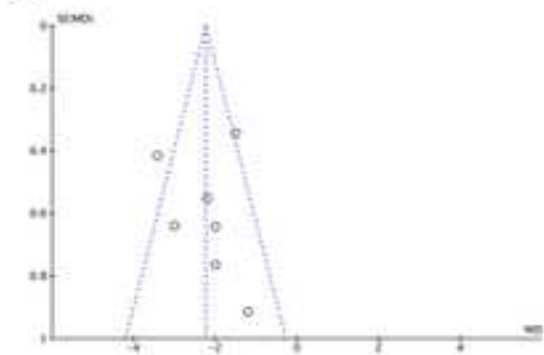

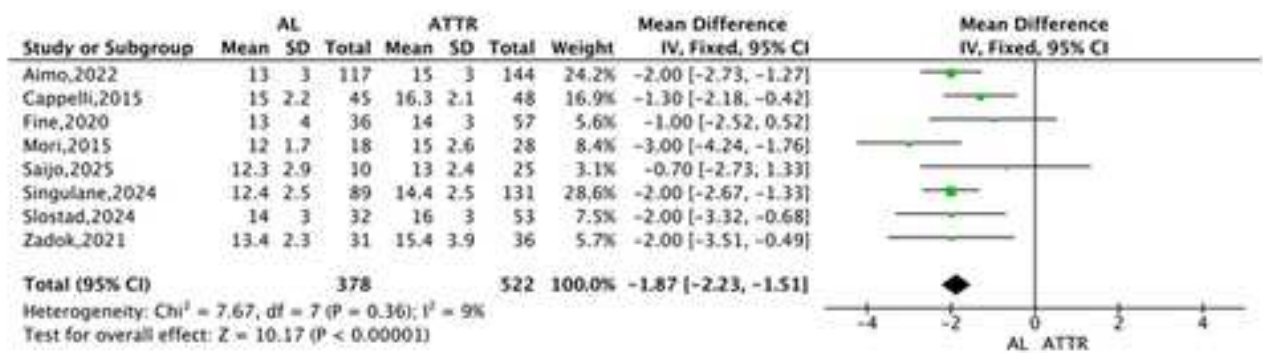

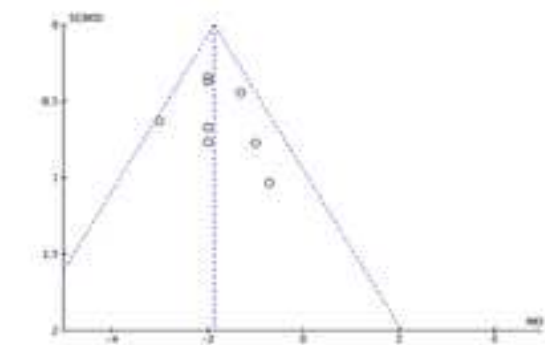

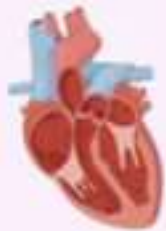
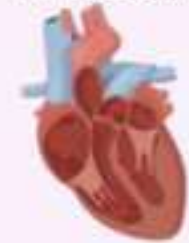